\documentclass[conference]{IEEEtran}
\IEEEoverridecommandlockouts

\usepackage{cite}
\usepackage{amsmath,amssymb,amsfonts}
\usepackage{graphicx}
\usepackage{textcomp}
\usepackage{xcolor}
\def\BibTeX{{\rm B\kern-.05em{\sc i\kern-.025em b}\kern-.08em
		T\kern-.1667em\lower.7ex\hbox{E}\kern-.125emX}}

\usepackage{amsmath,amsfonts,amsthm,amssymb}
\usepackage{mathrsfs}
\usepackage{dutchcal}
\usepackage{algorithm}
\usepackage{algorithmicx}
\usepackage{algpseudocode}
\usepackage{epsfig}
\floatname{algorithm}{Algorithm}

\theoremstyle{remark}
\newtheorem{theorem}{\hskip 1em Theorem}
\newtheorem{proposition}{\hskip 1em Proposition}

\newtheorem{lemma}{\hskip 1em Lemma}
\newtheorem{remark}{\hskip 1em Remark}

\newtheorem{definition}{\hskip 1em Definition}

\newtheorem{proof skecth}{Proof skecth}
\usepackage{array,caption}
\usepackage[mathscr]{eucal}
\usepackage[caption=false,font=LARGE,labelfont=rm,textfont=rm]{subfig}
\usepackage{textcomp}
\usepackage{stfloats}
\usepackage{url}
\usepackage{verbatim}
\usepackage{graphicx,float,balance}
\usepackage{multirow}
\usepackage{cite}
\usepackage{soul}
\usepackage{color,xcolor}
\begin{document}	
	\title{Reliability-Dependent Scaling Laws of Deterministic Identification over Binary Symmetric Channels
	}
	
	\author{
		\IEEEauthorblockN{Zhicheng Liu\IEEEauthorrefmark{1}, Liuquan Yao\IEEEauthorrefmark{2}, Guiying Yan\IEEEauthorrefmark{2}, Zhiming Ma\IEEEauthorrefmark{2} and Zechun Hu\IEEEauthorrefmark{1}}
		\IEEEauthorblockA{\IEEEauthorrefmark{1}%
			School of Mathematics, Sichuan University, Chengdu, China} 
		\IEEEauthorblockA{\IEEEauthorrefmark{2}%
			Academy of Mathematics and Systems Science, University of Chinese Academy and Sciences, Beijing, China}
		Email: lzc2025@scu.edu.cn, yaoliuquan20@mails.ucas.ac.cn, 
		yangy@amss.ac.cn, mazm@amt.ac.cn, zchu@scu.edu.cn
	}
	
	\maketitle

	\begin{abstract}
		In this paper, we study the asymptotic behavior of deterministic identification (DID) over binary symmetric channels (BSCs) under vanishing error constraints. By introducing a minimum error parameter, we characterize how different error-decay regimes affect the achievable DID rate. General achievability and converse bounds are derived, with explicit asymptotic characterizations in the large-deviation, moderate-deviation, and central-limit regimes. The achievability analysis combines coding-theoretic constructions with probabilistic concentration techniques, while the converse links statistical distinguishability to the minimum-distance structure of DID codes via total variation and Hamming-type bounds. Our results show that the asymptotic behavior of DID over BSCs is governed by a Hamming-shell concentration geometry of channel outputs, offering insights into the finite-blocklength behavior of deterministic identification over discrete-output channels.
	\end{abstract}

	
	\section{Introduction}

\IEEEPARstart{W}{ith} the rapid development of Internet-of-Things (IoT) systems~\cite{IoT_6G,IoT_survey}, autonomous driving, tactile internet, and large-scale machine-type communications, future wireless networks require ultra-reliable and low-latency connectivity among massive numbers of devices. In many applications, the goal is not full message reconstruction but event detection or verification, as in authentication, watermarking, control signaling, and vehicle-to-everything (V2X) systems~\cite{Secure_ID_wiretap,V2X_traffic,Millimeter_wave_MAS,Tactile_Internet,Robust_WHT_GFDM}. This motivates \emph{identification via channels} (ID), introduced by Ahlswede and Dueck~\cite{RID_Ahlswede}.

Unlike Shannon transmission~\cite{Shannon_communication}, where the receiver decodes the entire message, ID only requires deciding whether a given message was sent. Ahlswede and Dueck showed that randomized identification (RID) achieves doubly exponential scaling $\log\log N \sim nR$~\cite{RID_Ahlswede}, where $N$, $n$, and $R$ denote the number of identifiable messages, blocklength, and identification rate. Subsequent works developed RID via channel resolvability~\cite{Channel_Resolvability_Verdu}, converse bounds~\cite{Steinberg_RID_converse,RID_continous_alphabets}, information-spectrum methods~\cite{Han_Information_Spectrum,Hayashi_RID_second_order}, finite-blocklength analysis~\cite{Watanabe_Minimax_converse,Second_Order_RID_AWGN_Liu}, and explicit constructions~\cite{Verdu_explicit_optimal_CWC_ID}.

In many systems, however, encoder randomization is impractical, motivating \emph{deterministic identification} (DID). In contrast to RID, DID over discrete memoryless channels (DMCs) achieves only exponential scaling \(\log N \sim nR\). Early works by Ja’Ja’~\cite{JaJa_DID} and Ahlswede--Cai~\cite{Cai_DID} established the DID capacity for binary symmetric channels (BSCs) and DMCs.

The deterministic identification capacity of DMCs, including the BSC, was characterized in~\cite{Boche_DID_power_constraints}, and was subsequently placed within a broader geometric framework for finite-output channels in~\cite{Boche_DID_finite_output_superlinear}, and the AWGN case was resolved in~\cite{Pau_DID_Second_Order_AWGN}. Beyond DMCs, AWGN, Poisson, fading, and continuous-input Bernoulli channels may exhibit linearithmic message growth, $\log N=\Theta(n\log n)$~\cite{Secure_ID_Gaussian_Boche,Boche_DID_fading,Boche_DID_Possion,6G_Post_Shannon_Boche,Pau_DID_Bernoulli}. General rate--reliability bounds based on packing and covering numbers were recently established for discrete-output channels in~\cite{Pau_DID_Rate_Reliability} and extended to Gaussian channels in~\cite{Pau_DID_Rate_Reliability_Gaussian}. These studies reveal a fundamental distinction between continuous-alphabet channels and DMCs. For the former, stronger reliability requirements may reduce the identifiable message length from the linearithmic scale $\Theta(n\log n)$ to the linear scale $\Theta(n)$, whereas the BSC remains intrinsically in the linear regime and reliability affects only the attainable rate within that scale.

In this paper, we study the rate--reliability asymptotics of deterministic identification over BSCs under blocklength-dependent vanishing errors. By exploiting the Hamming-shell concentration geometry of BSC outputs, we derive explicit achievability and converse bounds across different error-decay regimes and characterize how the reliability requirement affects the convergence of the DID rate toward capacity. Our results can be viewed as an explicit BSC specialization of the general packing- and covering-based framework in~\cite{Pau_DID_Rate_Reliability}, yielding binary-entropy bounds with explicit channel- and reliability-dependent constants.

	\section{Problem Formulation}
	
	\subsection{Deterministic Identification Codes}
	\begin{definition}[DID Code]
		Consider a DMC \(W:\mathcal X\to\mathcal Y\), where $\mathcal X$ and $\mathcal Y$ denote the finite input and output alphabets, respectively.
		An $(n,N,\lambda_1,\lambda_2)$ DID code~\cite{Cai_DID} for $W$ is a collection
		\[
		\left\{(u_i,\mathcal D_i):i=1,2,\ldots,N\right\},
		\]
		where \(u_i\in\mathcal X^n\) is the codeword associated with message $i$, and
		\(\mathcal D_i\subseteq\mathcal Y^n\) is the corresponding decoding region.
		
		For each message $i$, define the type-I error probability
		\[
		P_{\text{I}}:=
		W^n(\mathcal D_i^{c}\mid u_i),
		\]
		and for any pair \(i\neq j\), define the type-II error probability
		\[
		P_{\text{II}}:=
		W^n(\mathcal D_i\mid u_j).
		\]
		
		An \((n,N,\lambda_1,\lambda_2)\) DID code is said to be achievable if
		\[
		P_{\text{I}}\le\lambda_1,
		\qquad
		P_{\text{II}}\le\lambda_2,
		\]
		for all \(i\neq j\).
	\end{definition}
	
	Consider the binary symmetric channel with crossover probability
	\(
	\delta\in(0,1/2),
	\)
	denoted by \(
	W=\mathrm{BSC}(\delta)
	\),
	whose output is given by
	\(
	Y=X\oplus N,
	\)
	where
	\(
	X\in\mathbb F_2^n
	\)
	and
	\(
	N\sim B(\delta),
	\)
	with
	\(
	B(\cdot)
	\)
	denoting the Bernoulli distribution and
	\(
	\oplus
	\)
	the addition over
	\(
	\mathbb F_2.
	\)
	Let
	\(
	N^*(W,n,\lambda_1,\lambda_2)
	\)
	denote the maximum number of identifiable messages achievable by an
	\(
	(n,N,\lambda_1,\lambda_2)
	\)
	DID code. The corresponding optimal DID rate is defined as
	\[
	R^*(W,n,\lambda_1,\lambda_2)
	:=
	\frac{
		\log N^*(W,n,\lambda_1,\lambda_2)
	}{n}.
	\]
	
	\begin{definition}[First-Order DID Capacity]
		The first-order DID capacity of the BSC is defined as
		\[
		C_{\mathrm{DID}}(W,\lambda_1,\lambda_2)
		:=
		\liminf_{n\to\infty}
		R^*(W,n,\lambda_1,\lambda_2).
		\]
	\end{definition}
	
	Ahlswede and Cai~\cite{Cai_DID}, with a later revisitation in~\cite{Boche_DID_finite_output_superlinear}, showed that for any fixed \(0<\lambda_1,\lambda_2<1\),
	\[
	C_{\mathrm{DID}}(W,\lambda_1,\lambda_2)=1,
	\]
	which strictly exceeds the Shannon transmission capacity
	\[
	C_{\mathrm{T}}(W)=1-h(\delta),
	\]
	where
	\[
	h(x):=-x\log x-(1-x)\log(1-x)
	\]
	is the binary entropy function. This highlights that DID over BSCs can support more identifiable messages than classical transmission, even without encoder randomization.
	
	Although the DID capacity characterizes the asymptotic growth rate, it does not capture finite-blocklength performance when the error probabilities vanish with \(n\). This motivates the study of second-order asymptotics to quantify the gap between finite-length DID rates and the asymptotic capacity.
	
	\subsection{Second-Order DID Asymptotics}
	To characterize finite-blocklength performance over BSCs, we introduce the following notion.
	
	\begin{definition}[Second-Order DID Asymptotics]
		Let \(\{\lambda_{1,n}\}_{n\ge1}\) and \(\{\lambda_{2,n}\}_{n\ge1}\) be type-I and type-II error constraints satisfying
		\begin{flalign*}
			P_{\mathrm I}\le \lambda_{1,n}, \qquad
			P_{\mathrm{II}}\le \lambda_{2,n}, \qquad \forall n.
		\end{flalign*}
		
		Define the optimal finite-blocklength DID rate as
		\[
		R_m(n)
		:=
		\frac1n
		\log N^*(W,n,\lambda_{1,n},\lambda_{2,n}).
		\]
		
		The second-order DID asymptotics describe the finite-blocklength gap
		\(C_{\mathrm{DID}}(W)-R_m(n),\)
		with emphasis on its dependence on the decay behavior of
		\(
		\eta_n:=\min\{\lambda_{1,n},\lambda_{2,n}\}
		\)
		as \(n\to\infty\).
	\end{definition}
	
	\subsection{Auxiliary Coding-Theoretic Results}
	We recall the classical GV and Hamming bounds in the binary Hamming space \(\mathbb F_2^n\), where \(\mathbb F_2\) denotes the binary field~\cite{Gilbert_bound,Varshamov_bound,Djordjevic_QEC}.
	
	Let
	\(
	A_2(n,d_m)
	\)
	denote the maximum size of a binary code
	\(
	\mathcal C_n\subseteq\mathbb F_2^n
	\)
	with minimum distance
	\(
	d_m\le n/2.
	\)
	
	\begin{lemma}\label{Lemma_Coding_Bounds}
		For any
		\(
		d_m\le n/2,
		\)
		\begin{flalign*}
			\frac{2^n}{
				\sum_{i=0}^{d_m-1}\binom ni
			}
			\le
			A_2(n,d_m)
			\le
			\frac{2^n}{
				\sum_{i=0}^{\lfloor (d_m-1)/2\rfloor}\binom ni
			},
		\end{flalign*}
		where \(\lfloor x\rfloor\) denotes the integer floor of \(x\), and the lower and upper bounds correspond to the GV and Hamming bounds.
	\end{lemma}
	
	Defining the normalized rate by
	\[
	R:=\frac1n\log A_2(n,d_m),
	\]
	some simple calculations yield that
	\begin{flalign}
		1-h\!\left(\frac{d_m}{n}\right)
		\le
		R
		\le
		1-h\!\left(\frac{d_m}{2n}\right).
		\label{eq:coding_bounds}
	\end{flalign}
	
	\section{Second-Order DID Asymptotics of the BSC}
	In this section, we establish finite-blocklength rate--reliability bounds for DID over the BSC. A target error of order $\exp(-\Theta(n^\alpha))$ induces achievable and necessary Hamming-separation scales of $n^{(1+\alpha)/2}$ and $n^\alpha$, respectively. Normalizing these distances and applying binary packing bounds yields the entropy terms in Theorem~\ref{Second_Order_asymptotics}.
\begin{theorem}[Rate--Reliability Bounds for the BSC]\label{Second_Order_asymptotics}
	Consider the binary symmetric channel $W=\operatorname{BSC}(\delta)$ with $0<\delta<\frac12$. For an $(n,N,\lambda_{1,n},\lambda_{2,n})$ DID code, define the reliability parameter
	\(
	\eta_n := \min\{\lambda_{1,n}, \lambda_{2,n}\}.
	\)
	
	We assume that $-\ln \eta_n$ scales polynomially with $n$, i.e.,
	\[
	-\ln \eta_n \asymp n^\alpha,\quad \alpha \in [0,1],
	\]
	where $\asymp$ denotes equality up to constant multiplicative factors for sufficiently large $n$.
	
	Then there exist positive constants $C_1, C_2$, depending only on $\delta$, such that the maximum rate $R_m(n)$ satisfies
	\begin{flalign*}
		1 - h\!\left(\frac{C_1}{n^{\frac{1-\alpha}{2}}}\right)
		\;\le\;
		R_m(n)
		\;\le\;
		1 - h\!\left(\frac{C_2}{n^{1-\alpha}}\right).
	\end{flalign*}
\end{theorem}

\begin{remark}
	Let $E(n):=n^{-1}\ln(1/\eta_n)\sim cn^{\alpha-1}$. For $0\leq \alpha<1$, Theorem~\ref{Second_Order_asymptotics} gives $R_m(n)\to1$, with achievability and converse backoffs of orders
	\[
	\sqrt{E(n)}\log\frac1{E(n)}
	\quad\text{and}\quad
	E(n)\log\frac1{E(n)},
	\]
	consistent with the general DMC bounds in~\cite{Pau_DID_Rate_Reliability}. For $\alpha=1$, a nonvanishing rate penalty remains. Thus, unlike continuous-alphabet channels, reliability over the BSC changes the rate value rather than the linear message-length scaling.
\end{remark}

	\noindent\emph{Geometric preparation.}
	To establish the achievability bound, we introduce the geometric structure induced by Hamming concentration of BSC outputs. For each \(u\in\mathbb F_2^n\), define the Hamming shell
	\[
	D_u(r)
	:=
	\{
	u\oplus v:
	v\in\mathbb F_2^n,\,
	|\sum_{i=1}^n v_i-n\delta|
	\le r
	\},
	\]
	which collects sequences whose Hamming weight concentrates around \(n\delta\) within fluctuation \(r\).
	
	Define the translated noise measure
	\[
	B_u(A)
	:=
	\mathbb P\!\left(
	N^n\in A\oplus u
	\right),
	\qquad u\in\mathbb F_2^n,\ A\subseteq\mathbb F_2^n,
	\]
	where \(N^n=(N_1,\ldots,N_n)\) with \(N_i\sim B(\delta)\). We write \(B_u\) for simplicity, and \(B_0\) for the all-zero input.
	
	With this notation, the type-I error reads
	\[
	P_{\mathrm I}
	=
	B_0(D_0^c)
	=
	\mathbb P\!\left(
	\left|
	\sum_{i=1}^n N_i-n\delta
	\right|
	>r
	\right).
	\]
	
	For two distinct \(u,v\in\mathscr C\), the type-II error is
	\[
	P_{\mathrm{II}}(u,v)
	=
	B_u(D_v)
	=
	\mathbb P(
	|\sum_{i=1}^n (u_i\oplus v_i\oplus N_i)-n\delta|
	\le r).
	\]
	
	Denote the minimum Hamming distance of some DID code \(\mathscr C\) by \(d_m\), and let \(w=(w_{1},\cdots,w_{n})\in\mathscr C\) attain it. By BSC symmetry and translation invariance, the worst-case type-II error is achieved by a nearest-neighbor pair:
	\[
	\max_{u\neq v} P_{\mathrm{II}}(u,v)
	=
	B_0(D_w)
	=
	\mathbb P\!\left(
	\left|
	\sum_{i=1}^n (N_i\oplus w_i)-n\delta
	\right|
	\le r
	\right).
	\]
	
The key point is that the coordinates where \(w_i=1\) and \(w_i=0\) contribute differently to the overlap probability, leading to a refined decomposition-based estimate.
	
	\begin{proposition}\label{type_II_err}
		Given an \((\lambda_1,\lambda_2,n)\)-DID code \(\mathscr C\) with minimum Hamming distance \(d_m\) and let \(M_i\sim B(1-\delta)\), then
		\begin{flalign*}
			\lambda_2
			\le\;&
			\mathbb P(
			\left|
			\sum_{i=1}^{n-d_m}N_i-(n-d_m)\delta
			\right|
			\ge r
			)
			\\
			&+
			\mathbb P(
			\sum_{i=1}^{d_m}M_i-d_m(1-\delta)
			\le
			2r-(1-2\delta)d_m).
		\end{flalign*}
	\end{proposition}
	
	This decomposition separates the type-II error into a concentration term outside the support of the minimum-distance vector and a fluctuation term along the nearest-neighbor direction. Consequently, the achievability analysis reduces to estimating concentration probabilities over Hamming shells, reflecting the underlying geometry of the second-order DID asymptotics. We now proceed with the proof, beginning with the achievability part.
	
	\noindent\emph{Proof of achievability.}
	The proof is divided according to the asymptotic behavior of
	\(
	-\ln\eta_n/n.
	\)
	We first consider the regime
	\(
	-\ln\eta_n=Ln
	\)
	for some \(L \ge 0\), then the moderate-deviation regime
	\(
	-\ln\eta_n = m n^\alpha
	\)
	with \(\alpha \in (0,1)\) and $m>0$, and finally the bounded-error regime
	\(
	\eta_n = O(1).
	\)
	These regimes correspond to fundamentally different reliability scalings and lead to distinct second-order asymptotic behaviors.
	
	\noindent\textbf{Case~1.} \emph{$-\ln\eta_n= Ln$ for some constant $L>0$.}
	
	In this regime, we show that both type-I and type-II errors can be controlled via appropriate choices of the shell radius \(r\) and the minimum distance \(d_m\). We first recall a standard large-deviation estimate for Bernoulli sums.
	\begin{lemma}[Large Deviation Bound {\cite[Theorem 2.7.7]{Durrett_PTE}}]
		Suppose
		\(
		\mathbb E\!\left(
		e^{\theta_1N}
		+
		e^{-\theta_2N}
		\right)
		<\infty
		\)
		for some \(\theta_1,\theta_2>0\), and the distribution of \(N\) is not a point mass at \(\delta\). Define
		\begin{flalign*}
			c_L(x)
			:=
			h(\delta)-h(\delta-x)
			-
			x\ln\frac{1-\delta}{\delta}
			>0,
			\qquad
			0<x\le\delta.
		\end{flalign*}
		Then for any \(0<a\le\delta\),
		\begin{flalign*}
			\mathbb P\!\left(
			\frac{
				\left|
				\sum_{i=1}^nN_i-n\delta
				\right|
			}{n}
			\ge a
			\right)
			\le
			2e^{-nc_L(a)},
			\qquad
			\forall n.
		\end{flalign*}
	\end{lemma}
	
	Since $-\ln\eta_n/n \to L$, we fix
	\(
	a_0 := c_L^{-1}\!\big(3(-\ln\eta_n/n)\big),\)
	and choose $r = c_l L n$, where $c_l>0$ depends only on $\delta$. For sufficiently large $n$, this ensures $r \ge a_0 n$.
	
	Applying the large deviation bound yields
	\begin{flalign*}
		&P_{\mathrm{I}}
		=
		\mathbb P\!\left(\left|\sum_{i=1}^n N_i - n\delta\right| > r\right)
		\le
		\mathbb P\!\left(\frac{1}{n}\left|\sum_{i=1}^n N_i - n\delta\right| \ge a_0\right) \\
		&\le 2e^{-n c_L(a_0)}\le \eta_n\le \lambda_{1,n},
	\end{flalign*}
	which verifies the type-I constraint  for sufficiently large $n$.
	
	\vspace{0.5em}
	
	We next control the type-II error. Let $d_m = t n$. For $M_i \sim B(1-\delta)$, a large deviation bound gives
	\begin{flalign*}
		\mathbb P\!\Big(
		\sum_{i=1}^{d_m} M_i - d_m(1-\delta)
		\le 2r - (1-2\delta)d_m
		\Big)
		\le 2e^{-t n c_L'(s)},
	\end{flalign*}
	where $s = (1-2\delta)t - 2c_l L$ and for any $0<x\le 1-\delta$
	\begin{flalign*} 
		c_L'(x) := h(1-\delta)-h(1-\delta-x) - x\ln\frac{\delta}{1-\delta}. 
	\end{flalign*}
	
	Since $c_L'(x) \ge c_l' x$ for some $c_l'>0$, this is further bounded by $\eta_n/2$ provided that
	\[
	t \ge \sqrt{
		\frac{
			2c_l'c_l L + (-\ln(\eta_n/4))/n
		}{
			c_l'(1-2\delta)
	}}.
	\]
	
	For the remaining term,
	\begin{flalign*}
		\mathbb P\!\Big(
		\left|\sum_{i=1}^{n-d_m} N_i - (n-d_m)\delta\right| \ge r
		\Big)
		\le 2e^{-(1-t)n c_L(c_l L/(1-t))}.
	\end{flalign*}
	
	If $t \le 1/2$, this term is also bounded by $\eta_n/2$ for sufficiently large $n$.	Therefore, choosing
	\[
	d_m = n \sqrt{
		\frac{
			2c_l'c_l L + (-\ln(\eta_n/4))/n
		}{
			c_l'(1-2\delta)
	}}
	\]
	guarantees that $P_{\mathrm{II}} \le \eta_n\le \lambda_{2,n}$.
	
Applying the lower bound in~(\ref{eq:coding_bounds}) with
\(
C_1=\sqrt{\frac{(2c_l'c_l+1)L}{c_l'(1-2\delta)}},
\)
yields the following achievable lower bound on the DID rate:
\begin{flalign*}
	R_{m}(n)
	\ge 1-h\!\left(\frac{d_m}{n}\right)= 1-h\!\left(
	\sqrt{
		\frac{
			2c_l'c_l L + \frac{-\ln(\eta_n/4)}{n}
		}{
			c_l'(1-2\delta)
		}
	}
	\right).
\end{flalign*}

In the regime $\alpha=1$, we have $-\ln \eta_n / n \to L$, and thus the above expression reduces to
\begin{flalign*}
	R_m(n)
	\geq 1 - h\!\left(
	\sqrt{\frac{(2c_l'c_l+1)L}{c_l'(1-2\delta)}}
	\right)= 1 - h(C_1),
\end{flalign*}
which completes the proof for this case.
	
	\noindent\textbf{Case~2.} \emph{$-\ln\eta_n = mn^\alpha$ for some $\alpha \in (0,1)$ and $m>0$.}
	
	We next consider the polynomial-decay regime, which lies between the central-limit and large-deviation scales. In this case, the bounds from Case~1 are no longer sufficient, and refined moderate-deviation estimates are needed. We therefore invoke the following result.
	\begin{lemma}[Moderate Deviation Bound {\cite[Theorem~2.2]{Moderate_Deviation_Reference}}]
		For any sequence $\{b_n\}$ satisfying
		\[
		\frac{b_n}{n} \to 0, \qquad \frac{b_n}{\sqrt{n}} \to \infty,
		\]
		there exist constants $c_M = \frac{a^2}{2\mathrm{Var}(N)} > 0$ and $n_a > 0$ such that:
		\begin{flalign*}
			\mathbb P(|\sum_{i=1}^n N_i - n\delta|/b_n \ge a)
			\le
			\exp(
			- c_M b_n^2/ n
			),\qquad \forall n>n_a.
		\end{flalign*}
	\end{lemma}
	
	We now return to the proof.
	
	Consider \(N\sim B(\delta)\). We choose
	\(
	b_n = n^{\frac{1+\alpha}{2}}
	\)
	and
	\(
	a=\sqrt{\delta(1-\delta)}
	\),
	and set
	\(
	r = a b_n = \sqrt{\delta(1-\delta)}\,n^{\frac{1+\alpha}{2}}.
	\)
	
	Since \(b_n/n\to 0\) and \(b_n/\sqrt n\to\infty\), the moderate-deviation conditions are satisfied. Moreover, \(c_M=\frac12\). Hence,
	\begin{flalign*}
		P_{\mathrm{I}}
		=
		\mathbb P\!\left(
		\left|\sum_{i=1}^n N_i - n\delta\right| \ge r
		\right)
		\le
		\exp\!\left(
		- c_M \frac{b_n^2}{n}
		\right)
		\le
		\eta_n
		\le
		\lambda_{1,n}.
	\end{flalign*}
	
	Let \(d_m = t n^{\frac{1+\alpha}{2}}.\) Then $r$ and $d_m$ operate on the same moderate-deviation scale. If \(t \le \frac{2\alpha}{(1+\alpha) a},\) then
	\(
	\frac{r}{n - d_m} \to 0,\) and \(\frac{r}{\sqrt{n - d_m}} \to \infty,\)
	so the first term in type-II error remains in the moderate-deviation regime.
	Consequently,
	\begin{flalign*}
		\mathbb P\!\left(
		\left|
		\sum_{i=1}^{n-d_m} N_i - (n-d_m)\delta
		\right|
		\ge r
		\right)
		\le
		\exp\!\left(
		- \frac{c_M b_n^2}{n-d_m}
		\right)\leq \frac{\eta_n}{2}.
	\end{flalign*}
	
	Moreover,
	\begin{flalign*}
		&\mathbb P\!\Big(
		\sum_{i=1}^{d_m} M_i - d_m(1-\delta)
		\le 2r - (1-2\delta)d_m
		\Big)\\
		\leq& \mathbb P\!\Big(
		\big|\sum_{i=1}^{d_m} M_i - d_m(1-\delta)\big|
		\ge ((1-2\delta)t-2a)n^{(\alpha+1)/2}
		\Big)\\
		\leq& \exp\Big(- c_M n^{\alpha}\Big)\leq \frac{\eta_n}{2}
	\end{flalign*}
	if
	\begin{flalign*}
		t\geq \frac{\sqrt{\frac{-2\delta(1-\delta)\ln \frac{\eta_n}{2}}{n^{\alpha}}}+2\sqrt{\delta(1-\delta)}}{1-2\delta}.
	\end{flalign*}
	
	Therefore, choosing
	\[
	d_m
	=
	\frac{
		\sqrt{
			-2\delta(1-\delta)\ln(\eta_n/2)/n^{\alpha}
		}
		+
		2\sqrt{\delta(1-\delta)}
	}{
		1-2\delta
	}
	\, n^{\frac{1+\alpha}{2}}
	\]
	leads to \(P_{\mathrm{II}} \le \eta_n\le\lambda_{2,n}\).
	
Using the lower bound in~(\ref{eq:coding_bounds}) with
\(
C_1=\frac{(2+\sqrt{2m})\sqrt{\delta(1-\delta)}}{1-2\delta},
\)
we obtain the following achievable rate.

By substituting the corresponding expressions into the bound, we obtain
\begin{flalign*}
	&R_m(n)
	\ge 1-h\!\left(\frac{d_m}{n}\right) \\
	&= 1-h\!\left(
	\frac{
		\sqrt{
			-2\delta(1-\delta)\ln(\eta_n/2)/n^{\alpha}
		}
		+
		2\sqrt{\delta(1-\delta)}
	}{
		1-2\delta
	}
	\, n^{\frac{\alpha-1}{2}}
	\right).
\end{flalign*}

Rearranging terms yields the scaling form
\begin{flalign*}
	R_m(n)
	\geq 1 - h\!\left(\frac{C_1}{n^{\frac{1-\alpha}{2}}}\right),
\end{flalign*}
which completes the proof for $\alpha\in(0,1)$.

	\noindent\textbf{Case~3.} \emph{$\eta_n = O(1)$.}
	
	We finally consider the central-limit regime, corresponding to
	\(
	-\ln\eta_n/n\to0.
	\)
	To capture the corresponding fluctuation behavior, we use the following central-limit estimate.
	\begin{lemma}[Central Limit Theorem {\cite[Theorem~3.6]{Chen_Stein}}]
		Suppose that
		\(
		\mathbb E|N-\delta|^3<\infty.
		\)
		Then there exists a constant
		\(
		c_C:=9.4\,\mathbb E|N-\delta|^3>0
		\)
		such that for any \(a>0\),
		\begin{flalign*}
			\mathbb P\!\left(
			\frac{\left|\sum_{i=1}^n N_i-n\delta\right|}{\sqrt n}
			\ge a
			\right)
			\le
			2\mathrm Q(a)
			+
			\frac{c_C}{\sqrt n},
		\end{flalign*}
		where
		\(
		\mathrm Q(x)
		=
		\int_x^\infty
		\frac{1}{\sqrt{2\pi}}e^{-t^2/2}\,\mathrm dt.
		\)
	\end{lemma}
	
	For the type-I error, choose
	\(
	a=\mathrm Q^{-1}(\eta_n/4)
	\)
	and
	\(
	r=a\sqrt n.
	\)
	Then, for sufficiently large \(n\),
	\begin{flalign*}
		P_{\mathrm I}
		=
		\mathbb P\!\left(
		\left|
		\sum_{i=1}^n N_i-n\delta
		\right|
		\ge r
		\right)
		\le
		2\mathrm Q(a)
		+
		\frac{c_C}{\sqrt n}
		\le
		\eta_n\le \lambda_{1,n}.
	\end{flalign*}
	
	Next, let
	\(
	d_m=t\sqrt n
	\)
	for some \(t>0\). Applying the same bound gives
	\begin{flalign*}
		\mathbb P\left(
		|\sum_{i=1}^{n-d_m} N_i-(n-d_m)\delta|
		\ge r
		\right)
		\le
		\frac{\eta_n}{2}.
	\end{flalign*}
	
	Moreover, for sufficiently large \(n\),
	\begin{flalign*}
		&\mathbb P\!\Big(
		\sum_{i=1}^{d_m} M_i-d_m(1-\delta)
		\le
		2r-(1-2\delta)d_m
		\Big)
		\\
		\le\;&
		\mathbb P\!\Big(
		\Big|
		\sum_{i=1}^{d_m} M_i-d_m(1-\delta)
		\Big|
		\ge
		((1-2\delta)t-2a)\sqrt n
		\Big)
		\\
		\le\;&
		2\mathrm Q((1-2\delta)t-2a)
		+
		\frac{c_C}{\sqrt{d_m}}
		\le
		\frac{\eta_n}{2},
	\end{flalign*}
	provided that
	\begin{flalign*}
		t
		\ge
		\frac{
			2\mathrm Q^{-1}(\eta_n/2)
			+
			\mathrm Q^{-1}(\eta_n/4)
		}{
			1-2\delta
		}.
	\end{flalign*}
	
	Therefore, choosing
	\[
	d_m
	=
	\frac{
		2\mathrm Q^{-1}(\eta_n/2)
		+
		\mathrm Q^{-1}(\eta_n/4)
	}{
		1-2\delta
	}
	\sqrt n
	\]
	ensures that
	\(
	P_{\mathrm{II}}\le \eta_n\leq\lambda_{2,n}.
	\)
	
	Since $\eta_n = O(1)$, without loss of generality we assume that $\eta_n = l$ for some constant $l>0$.
	
	The lower bound in~(\ref{eq:coding_bounds}) yields the following achievable rate by choosing \(C_1=\frac{2\mathrm{Q}^{-1}(l/2)+\mathrm{Q}^{-1}(l/4)}{1-2\delta}\):
	\begin{flalign*}
		R_{m}(n)
		&\ge 1-h\!\left(\frac{d_m}{n}\right) \\
		&= 1-h\!\left(
		\frac{
			2\mathrm{Q}^{-1}(\eta_n/2)
			+
			\mathrm{Q}^{-1}(\eta_n/4)
		}{
			(1-2\delta)\sqrt{n}
		}
		\right).
	\end{flalign*}
	
	Since $\eta_n=l$ is a constant in this regime, the above expression reduces to the canonical scaling form
	\begin{flalign*}
		R_m(n)
		\ge 1 - h\!\left(\frac{C_1}{\sqrt{n}}\right),
	\end{flalign*}
	which completes the proof for $\alpha=0$.

	Combining Cases~1--3 completes the proof of the achievability bound. We next turn to the converse analysis.\qed
	
	\noindent\emph{Proof of converse.}
	Unlike the achievability part, the converse is established by relating reliable identification to statistical distinguishability over the BSC. In particular, reliable decoding requires that the induced output distributions be sufficiently separated, which in turn imposes a lower bound on the minimum Hamming distance of the underlying codebook. The converse bound then follows by combining this distance constraint with the classical Hamming bound.
	
	We first establish an estimate on the total variation distance between shifted output distributions of the BSC.
	\begin{lemma}
		For any \(u\in\mathbb F_2^n\) with Hamming weight \(t\),
		\begin{flalign*}
			\frac12\|B_u-B_0\|_{\mathrm{TV}}
			\le
			1-2e^{-c_0 t},
		\end{flalign*}
		where \(c_0=c_0(\delta)>0\) depends only on \(\delta\), and
		\(
		\|P-Q\|_{\mathrm{TV}}
		=
		\sum_x |P(x)-Q(x)|
		\)
		denotes the total variation distance between probability measures \(P\) and \(Q\).
	\end{lemma}
	\begin{proof}
		By the definition of \(B_u\) and \(B_0\),
		\begin{flalign*}
			\frac12\|B_u-B_0\|_{\mathrm{TV}}
			=
			1
			-
			2\mathbb P\!\left(
			\sum_{i=1}^{t}N_i\ge \frac t2
			\right).
		\end{flalign*}
		
		Using the binomial probability formula,
		\begin{flalign*}
			\mathbb P\!\left(
			\sum_{i=1}^tN_i\ge \frac t2
			\right)
			\ge
			\mathbb P\!\left(
			\sum_{i=1}^tN_i=\frac t2
			\right)=
			{t\choose \frac{t}{2}}
			\delta^{\frac{t}{2}}(1-\delta)^{\frac{t}{2}}.
		\end{flalign*}
		
		Applying Stirling's approximation \cite[p.~45]{Wells_CuriousNumbers}, we obtain
		\begin{flalign*}
			{t\choose \frac{t}{2}}
			=
			\frac{t!}{((\frac{t}{2})!)^2}
			\approx
			\sqrt{\frac{2}{\pi t}}\,2^t,
		\end{flalign*}
		which implies
		\begin{flalign*}
			\mathbb P\!\left(
			\sum_{i=1}^tN_i\ge \frac t2
			\right)
			\ge
			\frac{c_1}{\sqrt t}
			\bigl(
			2\sqrt{\delta(1-\delta)}
			\bigr)^t
			\ge
			e^{-c_0 t},
		\end{flalign*}
		for some \(c_1>0\) and \(c_0=c_0(\delta)>0\). Substituting this estimate into the above identity completes the proof.
	\end{proof}
	
	The next result, established in \cite[Theorem~III.2]{Boche_DID_finite_output_superlinear}, shows that any DID code with sufficiently small type-I and type-II errors must induce well-separated output distributions.
	\begin{proposition}\label{Prop_DID_TV}
		For any DID code $\mathscr{C}$ with parameters \((\lambda_1,\lambda_2,n)\) over the BSC,
		\begin{flalign*}
			\frac12\|B_u-B_v\|_{\mathrm{TV}}
			\ge
			1-(\lambda_1+\lambda_2),
			\qquad
			\forall\,u,v\in \mathscr{C},\ u\neq v.
		\end{flalign*}
	\end{proposition}
	
	Combining the previous lemma with Proposition~\ref{Prop_DID_TV}, we obtain
	\(
	\lambda_1+\lambda_2 \ge 2e^{-c_0 t}.
	\)
	Since \(\lambda := \min\{\lambda_1,\lambda_2\}\) satisfies
	\(\lambda_1+\lambda_2 \ge 2\lambda\),
	it suffices to require
	\(
	\lambda \ge e^{-c_0 t},
	\)
	which is equivalent to
	\(
	t \ge -\ln \lambda / c_0(\delta).
	\)
	
	This immediately yields the following proposition.
	\begin{proposition}
		For any DID code \(\mathscr C\) with parameters \((\lambda_1,\lambda_2,n)\) over the BSC, the minimum distance \(d_C\) satisfies
		\begin{flalign*}
			d_C
			\ge
			-\ln\lambda/c_{0}(\delta),
		\end{flalign*}
		where
		\(
		\lambda=\min\{\lambda_1,\lambda_2\}
		\).
	\end{proposition}
	
	Combining the above minimum-distance constraint with the upper bound in~(\ref{eq:coding_bounds}) and setting \(\eta_n=\lambda\), we obtain
	\begin{flalign*}
		R_m(n)
		\le
		1-h(\frac{d_C}{2n})
		\le
		1-h(\frac{-\ln\eta_n}{2c_0(\delta)n})
		=1-h(\frac{n^{\alpha-1}}{2c_0(\delta)}),
	\end{flalign*}
	where the second inequality follows since \(h(x)\) is increasing on \((0,\frac{1}{2})\). Letting \(C_2=\frac{1}{2c_0(\delta)}\) completes the converse proof.
	
	Combining the achievability and converse results yields matching upper and lower bounds on the DID rate under the prescribed error constraint.\qed 
	
	The asymptotic behavior is therefore determined by the decay rate of $\eta_n$, leading to distinct scaling laws across the large-deviation, moderate-deviation, and central-limit regimes over BSCs. This highlights the fundamental interplay between reliability constraints and codebook geometry in characterizing the optimal DID rate.
	
	\section{Conclusion}
	In this paper, we studied the rate--reliability limits of deterministic identification over binary symmetric channels under blocklength-dependent vanishing error constraints. By combining coding-theoretic constructions with concentration arguments and minimum-distance converses, we derived explicit achievability and converse bounds across different reliability regimes. These results provide a BSC-specific realization of the general packing- and covering-based framework, with explicit binary-entropy expressions and channel-dependent constants. In particular, the optimal rate approaches the DID capacity of the BSC for subexponentially vanishing errors, although the convergence becomes slower as the reliability requirement increases, whereas exponentially vanishing errors result in a nonvanishing rate penalty. Future work includes tightening the gap between the achievability and converse bounds and extending the explicit analysis to broader classes of discrete memoryless channels.

	\section*{Acknowledgments}
	This work was supported by the National Key R\&D Program of China under Grant 2023YFA1009603. 
	
		%
			%
		
			%

		

	\end{document}